\documentstyle[preprint,aps]{revtex}
\tighten
\input{epsf.sty}
\preprint{CLNS 01/1738}

\font\blackboard=msbm10
\font\blackboards=msbm7 \font\blackboardss=msbm5
\newfam\black \textfont\black=\blackboard
\scriptfont\black=\blackboards \scriptscriptfont\black=\blackboardss

\def\be{\begin{equation}}
\def\ee{\end{equation}}
\def\ba{\begin{eqnarray}}
\def\ea{\end{eqnarray}}
\def\nn{\nonumber}
\def\Le{\Lambda_{\text{eff}}}

\def\li{L_i}

\def\ki{k_i}

\def\yi{y_i}

\def\th{$^{\text{th}}$ }
\def\hf{{1\over2}}

\begin{document}
\title{Probe Brane Dynamics and the Cosmological Constant}
\author{\'Eanna Flanagan\footnote{flanagan@spacenet.tn.cornell.edu},
Richard J. Hill\footnote{rjh@mail.lns.cornell.edu},
Nicholas Jones\footnote{nicholas-jones@cornell.edu},
S.-H. Henry Tye\footnote{tye@mail.lns.cornell.edu} and 
Ira Wasserman\footnote{ira@spacenet.tn.cornell.edu}}
\address{Laboratory for Nuclear Studies and 
Center for Radiophysics and Space Research \\
Cornell University \\
Ithaca, NY 14853}

\medskip
\date{\today}
\maketitle

\begin{abstract}

Recently a brane world perspective on the cosmological constant
and the hierarchy problems was presented. Here, we elaborate on 
some aspects of that particular scenario and discuss the stability 
of the stationary brane solution and the dynamics of a probe brane. 
Even though the brane is unstable 
under a small perturbation from its stationary position, such 
instability is harmless when the 4-D cosmological constant 
is very small, as is the case of our universe. One may also introduce 
radion stabilizing potentials in a more realistic scenario.

\end{abstract}

\section{Introduction}

The smallness of the cosmological constant is an outstanding fine-tuning 
problem. Recent observations \cite{new} indicate strongly that it has a 
positive value, implying that the small cosmological constant probably
has a dynamical origin (versus a zero value due to a symmetry reason). 
Many attempts addressing this fine-tuning
issue can be found in the literature. Here, we would like to elaborate on 
the recent brane world proposal of Ref\cite{TW,FJSTW}, which originates 
from a modification of the Randall-Sundrum model\cite{rs}.

In such a scenario\cite{FJSTW}, where the branes are stationary,
an exponentially small cosmological constant appears rather naturally. 
The scenario also easily incorporates
the solution to the hierarchy problem via warped geometry.
However, the general moving brane case is complicated by the 
coupling of the brane motions and the metric. 
This problem is tremendously simplified in the limit of zero brane 
tension and zero brane charge (but with a finite non-zero ratio). 
Since the visible brane tension must be smaller than the Planck brane 
tension, we are led to consider the visible brane as a probe brane.
In this approximation, the probe brane motion does not change the 
background metric, i.e., there is no back-reaction. 
It turns out that the resulting brane equation of motion is still 
rather non-trivial, but exact analytic solutions can be found. We shall 
discuss this case. In summary, the stationary branes are unstable: 
under small perturbations they will start moving. This conclusion is 
consistent with the result in other similar situations \cite{instability}. 
A stationary probe brane is stable if it is sitting at a minimum
of the warp factor. This corresponds to a particle horizon in our scenario.
Otherwise, an unstable mode is present and a perturbation away from 
the stationary position will grows like $e^{Ht}$, where $H$ is the Hubble 
constant.
So a stationary probe brane will start moving, eventually moving at 
the "speed of light" (an exact solution of the probe brane equation 
of motion). However, for a small effective 4-D cosmological constant 
(so a small $H$), the case we are interested in,
the growth of the perturbation is very slow.
Consider today's $H$, where $1/H$ is the cosmic size. A small 
perturabtion will only grow by one e-fold (say doubling) in the lifetime of 
the universe. That is, such an instability is harmless. 
We also discuss the implication of this result in the more general 
and realistic situations. In particular, the branes should be stabilized
with a radion potential\cite{goldwise}.

To check that there is no back reaction, and as an illustration, we 
consider explicitly a three $3$-brane model, where the visible brane is 
sitting between the Planck brane and a third brane, which is beyond the 
particle horizon. As the third brane and the Planck brane slowly 
move apart, the effective 4-D cosmological constant decreases. As the 
visible (probe) brane moves away from the Planck brane, the effective 
electroweak scale decreases, so there is a constraint on such motion 
as well. 

We also clarify the equivalence of two apparently different ways to 
obtain the effective 4-D cosmological constant given in Ref\cite{FJSTW}:
(i) via the determination of the Hubble constant $H$ and (ii) via the 
4-D effective action. In the latter approach, besides the contributions 
from the brane tensions and the bulk cosmological constants,
the extra dimension curvature also contributes. In the large brane 
separations case, the various terms almost cancel, reproducing the 
exponentially small 4-D effective cosmological constant found by the 
other approach.

\section{The Model} 

Consider the action $S$ consisting of a bulk action  
containing 5-D gravity, with metric $ {g}_{ {a} {b}}$ ($a=0,1,2,3,5$), 
and a set of worldvolume actions of $3$-branes with
dynamical coordinate fields ${X}_{n}^{{a}}$.
This means the location of the $n$\th brane in the bulk is given by
the embedding functions $x^a = X^a_n(\xi_n^\mu)$, and
the $n$\th brane possesses a brane metric
$\gamma_n^{\mu\nu}$ which is a function of the brane
coordinates $\xi_n^\mu$ ($\mu=0,1,2,3$). The $n$\th brane couples 
(with charge $e_{ni}$) to the
$4$-form potentials ${A}_{i} \equiv A_{i a_1\cdots a_{4}}(X_n)$ 
(with $5$-form field strength ${F}_{(i)}=d{A}_{i}$).
The action for such a system is\cite{polch,fst3,FJSTW}
\ba
\label{Action}
&& S=\int d^5x\sqrt{\vert g\vert}\left[\frac{1}{2\kappa^2}R-\Lambda-
\frac{1}{2\cdot 5!}F^2_{\left(i\right)}\right] \nonumber \\
&& +\sum_n\left\{-\frac12\sigma_n\int d^{4}\xi_n\sqrt{\vert\gamma_n\vert}
\left[\gamma_n^{\mu\nu}
\partial_{\mu}X^a\partial_{\nu}X^b g_{ab}\left(X_n\right)-2\right]
\right. \nonumber \\
&&-  \left.\frac{e_{ni}}{4!}\int d^{4}\xi_nA_{i 
a_1\cdots a_{4}}\left(X_n\right)
\partial_{\mu_1}X^{a_1}\cdots\partial_{\mu_{4}}X^{a_{4}}\epsilon^{\mu_1
\cdots\mu_{4}} \right\}
\ea
where $\sigma_n$ and $e_{ni}$ are the tension and the charges (coupling to
the $4$-form potential $A_i$ of the 
$n$\th brane), respectively. In the action $S$, the sum over $i$ is implicit.
The tensor density is totally antisymmetric with $\epsilon^{0123}=1$.
Varying the action $S$ with respect to the metric $ {g}_{ {a} {b}}$, 
the $4$-form potential ${A_{i}}$, the brane metric $\gamma_{\mu \nu}$ and 
the brane coordinates $X^a$, we have, respectively,
the 5-D Einstein equation, the $4$-form field equation, the induced metric 
relation and the brane equation of motion.
In particular, the brane equation of motion for the $n$\th brane is given by
\ba
\label{braneeq}
&& \sigma_n \sqrt{\vert\gamma_n\vert}\left[\nabla_{\mu}\nabla^{\mu}X_n^a\
+\Gamma^a_{bc}\partial_{\mu}X^b_n\partial_{\nu}X^c_n
\gamma_n^{\mu\nu}\right]+\nonumber \\
&& \frac{e_{ni}}{\left(4\right)!}F^{ia}_{b_1\cdots b_{4}}\partial_{\mu_1}
X_n^{b_1}\cdots
\partial_{\mu_{4}}X_n^{b_{4}}\epsilon^{\mu_1\cdots\mu_{4}}=0.
\ea

Solving the $4$-form field equations, we see that the background 
field strengths are constants to be determined by the Einstein equation.
For parallel stationary branes with an appropriate number of $4$-form 
potentials, the complete set of equations have been solved in Ref\cite{FJSTW}. 
Here we shall review some of the key features of the more general solution 
by considering the specific case of three parallel stationary $3$-branes 
with the 5\th dimension uncompactified. The two brane compactified case
and the two brane orbifold case discussed in Ref\cite{TW,FJSTW} can be 
easily obtained from this case (by identifying two of the branes, and 
by orbifolding, respectively). 
The above brane equation of motion also has moving brane solutions. 
They are rather complicated and will be discussed elsewhere. 
In this note, we shall
only consider the brane motion for a brane with zero brane tension and zero 
brane charge (but with a non-zero charge to tension ratio). 

\section{Motion of the Visible Brane as a Probe Brane}

The Einstein equations were solved in Ref\cite{TW,FJSTW,kaloper}.
Starting with the metric ansatz
\be
\label{metric}
  ds^2 = dy^2+A(y)[-dt^2+e^{2Ht}\delta_{ij}dx^idx^j]
\ee
the Einstein equations for the bulk are
\ba
\label{eq:einstein}
A^{\prime\prime}&=&2H^2+4k^2A \nonumber \\
{(A^\prime)^2\over A}&=&4H^2+4k^2A
\ea
This yields the general solution for the bulk
\be
\label{metricA}
  A(y) = {H^2\sinh^2[\ki(y-\yi)]\over\ki^2},
\ee
in which $\ki\equiv\sqrt{\kappa^2\Lambda_i/6}$ and $y_i$ are
in general integration constants.
Here, $-\Lambda_i$ are the bulk cosmological 
constants that depend on the constant $5$-form field strengths,
\be
\label{lambda1}
-\Lambda_i =-\Lambda+
\frac12\sum_j\left(e_{Bj}+e_{0j}+e_{1j}+...+e_{(i-1)j}\right)^2.
\label{manycharges}
\ee
Here we require $\Lambda_i>0$ for Anti-deSitter spaces between branes.
This requires $-\Lambda$ to be negative enough.
Depending on the charges of the branes, this will allow some number
of bulk cosmological constants to be determined by the Einstein
equation, or equivalently, by the boundary conditions at the branes 
(that is, the jump conditions \cite{Israel}). There,
the warp factor $A(y)$ is continuous across the branes while 
$A^\prime (y)$ has a jump at the brane as a function of the brane tension.

Let us now consider the brane equation of motion for branes that may not be 
stationary. In general, brane motions and the metric are coupled in a way 
that do not allow a simple explicit analytic solution. 
To see what is going on, one may consider slow brane motion, or 
alternatively, the situation where the particular brane tension is 
negligible. Here we shall consider the latter case, i.e., a probe (or test)
brane. In particular, we may consider the case where the visible brane 
is the probe brane. As we shall see more explicitly, as the probe 
brane tension $\sigma_1 \rightarrow 0$, both $H$ and $A(y)$ are 
independent of the position of the visible brane. In this limit,
the brane equation of motion is still non-trivial, but it turns out that
explicit analytic solutions can be found. This is clearly easier than 
the full dynamics. Understanding the probe brane motion can give 
insight into the full dynamics. We could also study the problem 
perturbatively, treating the probe brane tension as a small parameter. 

In the above scenario, we shall consider a set of branes with non-zero 
brane tensions and charges, so that $A(y)$ is solved.
Next we introduce the visible brane, namely the $\sigma_1$ brane, 
which is coupled to only one $4$-form potential, say $A_1$.
Now we shall take both this charge, namely $e_{11}$, and tension $\sigma_0$
to zero, namely $e_{11} \rightarrow 0$ and $\sigma_1 \rightarrow 0$,
with a finite nonzero charge-to-tension ratio, $e_{11}/\sigma_1=Q$.
The brane equation of motion 
(\ref{braneeq}) for such a probe brane (at $y$) reduces to
\be
0=\ddot y+2A^\prime-{QE_1(A-\dot y^2)^{3/2}\over\sqrt{A}}
+3H\dot y-{5A^\prime\over 2A}\dot y^2-{3H\over A}\dot y^3,
\label{eom}
\ee
in the presence of the background metric (\ref{metric})
and a constant background 5-form field strength $E_1$. 
Since all the probe brane tension and charges are zero, there 
is no back-reaction. As the probe brane moves, both $H$ and the metric 
$A(y)$ remain unchanged,
as long as the other branes (ones that have non-zero tensions and charges)
are stationary.

There are two exact solutions to this equation : \\
$\bullet$ One corresponds to the equilibrium point
\be
QE_1={2A^\prime(y)\over A(y)} \qquad \dot y=0,
\ee
presuming it exists. In our set up, the background field strength $E_1$ is 
an integration constant, so, for any position $y$ for the brane, we 
may fix the value $E_1$ to satisfy the above equation. 
This simply reproduces the stationary solution we have been discussing.
However, once $E_1$ is fixed, its value (and so the bulk cosmological 
constant) will stay constant, at least classically.  \\
$\bullet$ The other solution is
\be
\dot y=\pm\sqrt{A(y)}.
\ee
This solution corresponds to `motion at the speed of light'. 

These are the two exact analytic solutions: one is stationary at the
equilibrium point, and the other has the brane moving at `light'
speed. The equation of motion is an initial value
problem, and has an infinite number of solutions.
We have found that a stationary brane is a solution. Suppose 
this is the initial situation. Let us check the stability of this 
stationary solution by asking what happens if the probe brane is 
perturbed away from equilibrium. To find out,
consider small perturbations around the equilibrium point, 
$y=y_0+\delta y(t)$.
Linearizing Eq.~(\ref{eom}) implies
\ba
\label{pert}
0&=&\delta\ddot y+3H\delta\dot y+[2A^{\prime\prime}_0-QE_0A^\prime_0]\delta y
\nonumber\\
&=&\delta\ddot y+3H\delta\dot y+2\biggl[A^{\prime\prime}_0-{(A^\prime_0)^2
\over A_0}\biggr]\delta y,
\ea
where the subscripts $0$ mean evaluate the derivatives at $y=y_0$, and
the second form of the equation uses the equilibrium relation to eliminate
$QE_1$. 
Using the Einstein equation (\ref{eq:einstein}),
the equation for linear perturbations can be reduced to
\be
\label{tachyon}
0=\delta\ddot y+3H\delta\dot y-4H^2\delta y.
\ee
This is the equation of motion for the radion mode.
This equations has solutions of the form $\delta y\propto e^{\sigma t}$ with
\be
0=\sigma^2+3H\sigma-4H^2=(\sigma-H)(\sigma+4H),
\ee
implying two roots, $\sigma=-4H$, which is stable, and $\sigma=H$, which is
unstable. Thus, the equilibrium point is unstable, and the probe brane will
tend to move away from it if perturbed at all. This instability property 
is consistent with the conclusion of similar analyses in similar 
scenarios \cite{instability}. What happens is that a probe brane perturbed 
away from equilibrium accelerates until it approaches the motion at speed 
of `light', and, according to that solution, eventually wants to settle at 
a zero of $A(y)$, either by passing through
the Planck (the $\sigma_0$) brane or not.  

The timescale for the brane
to move is $\sim H^{-1}$, which is long compared to $\sim k^{-1}$ if $H\ll k$
but the instability may still grow on a timescale relevant for cosmology. For
example, the value of $A$ on the brane changes by a fractional amount
$(A_0^\prime/A_0)\delta y_+(0)e^{Ht}$, where the amplitude of the growing mode
at $t=0$ is $\delta y_+(0)$. The change is substantial after a time of order
$H^{-1}\ln[A_0/A^\prime_0\vert\delta y(0)\vert]$, which would imply, for
example, considerable cosmological evolution of the Higgs mass scale.

The time evolution of the displacement should proceed exponentially at first,
according the linear instability, and then asymptote toward motion at the
speed of light.  In this limit, the rate of change of $A$ at the probe brane
is governed by
\be
{\dot A\over HA}=\pm 2\sqrt{1+{k^2A\over H^2}},
\ee
and is always large on a cosmological timescale (and is actually large on
the timescale $k^{-1}$ generally).

Let us consider what happens when $Q\equiv 0$, i.e.
when the charge-to-tension
ratio vanishes exactly. In this case, eq.
(\ref{eom}) becomes
\be
0=\ddot y+2A^\prime
+3H\dot y-{5A^\prime\over 2A}\dot y^2-{3H\over
A}\dot y^3,
\label{eomo}
\ee
and the equilibrium point(s) are the extrema of $A$,
i.e., $A^\prime=0$ at
$y=y_0$. Linearizing around this point implies the
perturbation equation
(which can also be read off from the first line of
Eq. (\ref{pert}))
\be
0=\delta\ddot y+3H\delta\dot
y+2A_0^{\prime\prime}\delta y,
\ee
which implies decaying oscillations when
$A_0^{\prime\prime}>0$.
Thus, the probe brane can sit stably at the minimum
of $A$ when it has
$Q\equiv 0$. For the AdS metric, $A(y)=0$ at the
minimum, but in non-AdS,
the minimum value of $A(y)\neq 0$ in general.

Eq(\ref{tachyon}) shows that this model is unstable 
(with tachyonic $m^2 = -4H^2$), but also suggests
that this instability could be generic. We shall discuss stability
calculations that support this supposition elsewhere.
This instability in general will cause other 
problems for the model. One may fix this problem by introducing a radion 
stabilizing potential for this mode \cite{goldwise}. For example, we 
may consider a potential with its minimum at $y=L$, 
\be
V(y)= -\Lambda + C(y-L)^2 + ...
\ee
where $C>0$ so the probe brane is stable at $y=L$. Here $-\Lambda$ is 
simply the term that contributes to the bulk cosmological constant 
(\ref{lambda1}).
In the scenario with more than one probe brane, the various branes 
tend to fall into the same position, causing brane collisions that 
originate the big bang. Inter-brane interactions may lead to small
brane separations, so our visible brane may sit close to a hidden 
brane that contains some or all of the dark matter we see today.
The situation in a more realistic scenario will be discussed 
elsewhere.

The above toy model does suggest a number of interesting scenarios.
For example, suppose we have two probe branes. Typically they will 
collide when moving towards the stable point $y=L$.
In this scenario, the collision is expected to release energy into
radiation, and the two branes may fuse to form the visible brane after
the collision \cite{gia}. The resulting visible brane may be relatively
stationary.

\section{3 Branes at Large Separation}

Let us consider a specific scenario, namely a model with three parallel
$3$-branes and see explicitly how the visible brane becomes a probe 
brane when its brane tension and brane charge vanish, that is, the metric 
dependence on its position drops out.
We now calculate the Hubble constant directly from the 5-D gravitational 
equations of motion. Since the only source for the bulk stress energy tensor 
is the 5-D cosmological constant, we have Anti-deSitter (AdS) bulk spacetime 
and we assume the branes lie along deSitter (dS) foliations. 
For the case of three branes, with brane tensions $\sigma_i$ and 
brane charges $e_{ij}$, $i=0,1,2$, there are 2 background field strengths 
$e_{Bj}$, $j=1,2$, that are integration constants to be determined.
Here we require $\Lambda_i>0$ for AdS spaces between branes. 
This requires $-\Lambda$ to be negative enough.
Depending on the charges of the branes, this will allow some number
of bulk cosmological constants to be determined by the Einstein
equation, or equivalently, by the boundary conditions at the branes.
The model is shown in Figure 1, where we use the rescaled brane tension 
$q_i\equiv \kappa^2\sigma_i/3$
and $k_i^2\equiv \kappa^2\Lambda_i/6$.

\newcounter{figcount}
\setcounter{figcount}{1}
\begin{center}
  \epsfbox{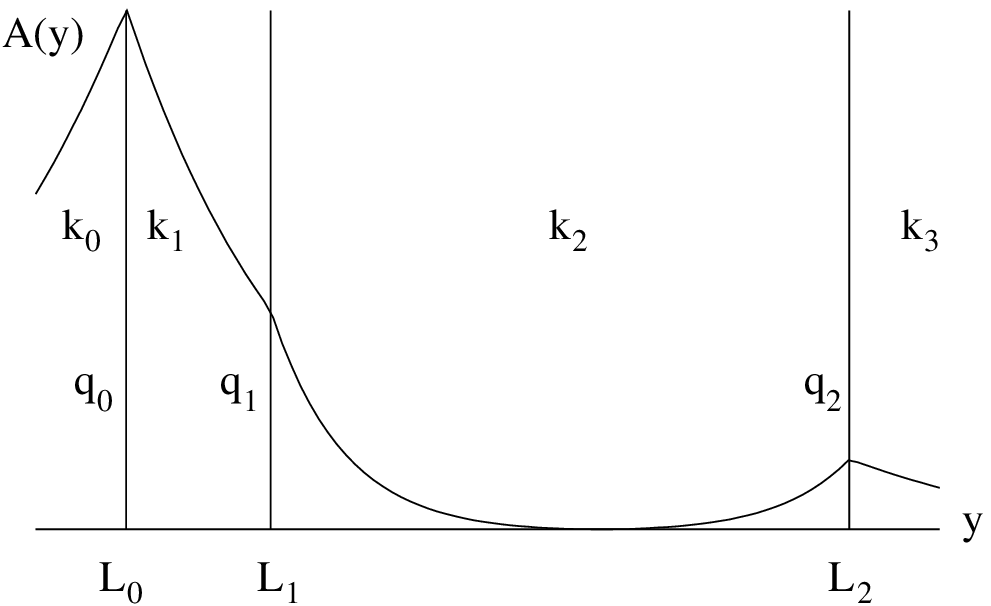}
  \parbox{12cm}{\vspace{1cm}
    FIG. \thefigcount \hspace{2pt} Schematic arrangement of the branes and 
    the warp factor $A(y)$.}
\end{center}

In this model, the above probe brane scenario corresponds to the 
situation where only the $\sigma_0$ and the $\sigma_2$ branes 
couple to the $4$-form potential $A_2$. There, the $\sigma_0$ brane 
is the Planck brane while
the $\sigma_1$ brane is the visible brane.
We shall also take both the visible brane charge, namely $e_{11}$, and 
its tension to zero, keeping the charge-to-tension ratio, 
$Q=e_{11}/\sigma_1$ fixed to a finite non-zero value. Before going to 
this limit, let us first consider the general case.

Integrating the $G_{00}$ equation across the branes 
leads to the Israel jump condition\cite{Israel}:
\ba
\lim_{y\to\li+}\biggl({A^\prime\over A}\biggr)
-\lim_{y\to\li-}\biggl({A^\prime\over A}\biggr)
&=&-{2\kappa^2\sigma_i\over 3}
\equiv -2q_i,
\label{Israel}
\ea
in addition to which, the metric and hence the scale factor, $A(y)$, is 
required to be continuous across the branes.

With a 5\th dimension of infinite extent, and three parallel 
3-branes located at $y=L_0,L_1,L_2$, the jump and continuity conditions 
become
\ba\nn
  && q_0 = k_0\coth[k_0(L_0-y_0)]-k_1\coth[k_1(L_0-y_1)],\\\nn
  && q_1 = k_1\coth[k_1(L_1-y_1)]-k_2\coth[k_2(L_1-y_2)],\\\nn
  && q_2 = k_2\coth[k_2(L_2-y_2)]-k_3\coth[k_3(L_2-y_3)],\\\nn
  && k_0^2(\coth^2[k_0(L_0-y_0)]-1)=k_1^2(\coth^2[k_1(L_0-y_1)]-1),\\\nn
  && k_1^2(\coth^2[k_1(L_1-y_1)]-1)=k_2^2(\coth^2[k_2(L_1-y_2)]-1),\\
  && k_2^2(\coth^2[k_2(L_2-y_2)]-1)=k_3^2(\coth^2[k_3(L_2-y_3)]-1).
  \label{jump_cont}
\ea
This is the setup of the problem. In this case, we like to consider 
two $4$-form potentials. So, of the four $k_j$'s, two of them, or two 
combinations of them are to be treated as integration constants.
Now we can solve for $H$ and $G_N$  and so $\Le$. Since we are interested 
mainly in the small $\Le$ case, we shall solve this set of equations in 
the limit of large $k_2(L_2-L_1)$ and 
$k_1(L_1-L_0)$.  Writing
\ba\nn
  t_1\equiv\tanh[k_1(L_1-L_0)]&\simeq&1-\epsilon_1,\quad\text{where }
    \epsilon_1\equiv2e^{-2k_1(L_1-L_0)},\\\nn
  t_2\equiv\tanh[k_2(L_2-L_1)]&\simeq&1-\epsilon_2,\quad\text{where }
    \epsilon_2\equiv2e^{-2k_2(L_2-L_1)},\\\nn
  \coth[k_0(L_0-y_0)]&\simeq&1+\eta_0,\\
  \coth[k_1(L_1-y_1)]&\simeq&-1-\eta_1.
\label{t1t2}
\ea
in which $\eta_{\,0}$ and $\eta_{\,1}$ are expected to be exponentially small 
in the interbrane separations.  To $\mathcal{O}$$(\epsilon^0)$, the jump 
conditions give 
\ba\nn
  q_0 &=& k_0+k_1,\\
  q_1 &=& k_2-k_1,
\ea
and to $\mathcal{O}$$(\epsilon^1)$, we have for $\eta_{\,0,1}$
\ba\nn
  &&\eta_{\,0}={1\over2}\left({q_0+q_1-k_0\over k_0}\right)^2
    \left[{\coth[k_2(L_2-y_2)]+1\over\coth[k_2(L_2-y_2)]-1}\right]
    \epsilon_1\epsilon_2,\\
  &&\eta_{\,1}=\left({q_0+q_1-k_0\over q_0-k_0}\right)^2
    \left[{\coth[k_2(L_2-y_2)]+1\over\coth[k_2(L_2-y_2)]-1}\right]
    \epsilon_2.\label{etas}
\ea
The jump and continuity conditions (\ref{jump_cont}) can be used to write
\be
  \coth[k_2(L_2-y_2)]={q_2^2+k_2^2-k_3^2\over2q_2k_2}.
\ee
Given that this expression does not approach unity, the system can be 
solved as
\ba\nn
  H_0^2\equiv {H^2\over A(L_0)} &=& 2k_0^2\eta_{\,0}
    =(q_0+q_1-k_0)^2\left[{(q_0+q_1+q_2-k_0)^2-k_3^2
    \over(q_0+q_1-q_2-k_0)^2-k_3^2}\right]\epsilon_1\epsilon_2,\\
  H_1^2\equiv {H^2\over A(L_1)} &=& 2k_1^2\eta_{\,1}
    =(q_0+q_1-k_0)^2\left[{(q_0+q_1+q_2-k_0)^2-k_3^2
    \over(q_0+q_1-q_2-k_0)^2-k_3^2}\right]\epsilon_2,\label{H_effs}
\ea
where $\epsilon_1$ and $\epsilon_2$ are given in Eq(\ref{t1t2}).
Here, $k_0$ and $k_3$ are treated as undetermined by the boundary and jump 
conditions. For given brane charges, $k_0$ and $k_3$ are related to
$k_1$ and $k_2$.

The important behavior of this solution lies in the exponentially small 
terms.  There is an exponentially large hierarchy between mass scales on 
the branes at $L_0$ and $L_1$, such that, as in Ref.~\cite{FJSTW} (also see
section~\ref{4da}),
\be
  \left({M_{Higgs}\over M_{Pl}}\right)^2 
    \sim\frac{A(L_1)}{\displaystyle \int dyA(y)}
    \simeq\frac{A(L_1)}{A(L_0)}\simeq e^{-2k_1(L_1-L_0)},
\ee
giving Planck masses exponentially greater than Higgs' masses with the 
exponent related to the distance scale between the Planck and visible 
brane.  Similarly, the exponential scale for the smallness of the 
cosmological constant with respect to the Planck mass arises from the 
distance between the $L_1$ and $L_2$ branes.  For large brane separations, 
the cosmological constant behaves like
\be
  {\Le\over M_{Pl}^4}\simeq{H_1^2\over M_{Pl}^2}
    \simeq{\epsilon_2A(L_1)\over A(L_0)}
    \simeq e^{-2[k_2(L_2-L_1)+k_1(L_1-L_0)]}
\ee
This behavior is generic of these such models, with the two different 
length scales in the 5D spacetime leading to the two different exponentially 
large physical scale diffenences on the visible brane.

This result (\ref{H_effs}) can be applied to a variety of scenarios.
Suppose that the $q_0$ and $q_2$ branes have no brane charges, i.e., 
$e_{0j}=e_{2j}=0$, $j=1,2$. Then $k_0 = k_1$ and $k_3 = k_2$, and 
(\ref{H_effs}) becomes
\ba\nn
  H_0^2 &=& 4\left({q_0\over2}+q_1\right)^2\left[{q_2+2q_1+q_0
    \over q_2-2q_1-q_0}\right]e^{-q_0(L_1-L_0)-(q_0+2q_1)(L_2-L_1)},\\
  H_1^2 &=& 2\left({q_0\over2}+q_1\right)^2\left[{q_2+2q_1+q_0
    \over q_2-2q_1-q_0}\right]e^{-(q_0+2q_1)(L_2-L_1)}.
\ea

We may also apply the result (\ref{H_effs}) to a model with a compactified 
5\th dimension. Here, $\coth[k_2(L_2-y_2)]\to1$ in (\ref{etas}).  
and we identify the branes at $L_0$ and $L_2$, such that
$q_2\equiv q_0$, $e_{0j} \equiv e_{2j}$ ($j=1,2$), $k_0 \equiv k_2$ and 
$k_3 \equiv k_1$. Because the 5\th dimension is compactified, charge 
conservation requires that $e_{0j} + e_{1j}=0$, $j=1,2$. Writing
\ba\nn
  \coth[k_2(L_2-y_2)]\equiv1+\eta_{\,2},
\ea
we can solve for $\eta_{\,2}$ using (\ref{jump_cont}) to be 
\ba\nn
  \eta_{\,2} = \sqrt{\epsilon_1\epsilon_2},
\ea
which when substituted into (\ref{etas}) gives the following expressions for 
$H_{0,1}^2$:
\ba\nn
  H_0^2 &=& (q_0+q_1)^2e^{-\hf[(q_0+q_1)(L_2-L_1)+(q_0-q_1)(L_1-L_0)]},\\
  H_1^2 &=& (q_0+q_1)^2e^{-\hf[(q_0+q_1)(L_2-L_1)-(q_0-q_1)(L_1-L_0)]}.
\ea
The result is simply the compactified model with two branes 
discussed in Ref\cite{FJSTW}. The probe brane case simply requires
both $e_{11} \rightarrow 0$, and $\sigma_1 \rightarrow 0$, while
keeping $Q=e_{11}/\sigma_1$ fixed to a finite non-zero value.

Let us consider the overall picture of the three brane model,
where the visible brane is
sitting between the Planck brane and a third brane, which is beyond the
particle horizon. As the third brane and the Planck brane slowly
move apart, the effective 4-D cosmological constant decreases. 
We can use this motion to accommodate the quintessence picture.
As the visible brane moves away from the Planck brane, the effective
electroweak scale decreases, so there is a constraint on such motion.
An analysis of this scenario will be interesting.

\section{$\Le$ from 4-D Effective Action}\label{4da}

It is instructive to consider the derivation of $\Le$ starting from 
the 5-D action ${\bf S}^{(5)}$:
\be
{\bf S}^{\left(5\right)}=\int dyd^4x \sqrt{-g}\left[\frac{R^{\left(5\right)}}
{2\kappa^2} + \Lambda + \sum_i \delta (y-L_i)\sqrt{g^{55}}
(-\sigma_i + {\bf L}_i)\right]+ 
\int_{\Sigma}\frac{K\sqrt{-\gamma}}{\kappa^2}d^4x,
\ee
with ${\bf L}_i$ being the $i$\th brane mode Lagrangian. The last term 
is the Hartle-Hawking boundary term.
In obtaining the same results as found by algebraic means, 
the sources of the various cancelling contributions, and their
geometric interpretations, will become clear.

We use the 5-D metric ansatz,
\be
ds^2=A\left(y\right)\hat \gamma_{\mu\nu}\left(x\right)dx^{\mu}dx^{\nu}+dy^2,
\ee
and the decomposition of the 5-D Ricci tensor, 
$R_{\mu \nu}^{\left(5\right)}(g_{ab})$,  into its 4-D counterpart, 
$R_{\mu \nu}^{\left(4\right)}(\hat\gamma_{\mu \nu})$,
and the extrinsic curvature $K_{\mu\nu}$ ($K\equiv g^{\mu \nu} K_{\mu \nu}$):
\be
R_{\mu \nu}^{\left(5\right)}= R_{\mu \nu}^{\left(4\right)}-
g^{55}\partial_yK_{\mu\nu}-g^{55}K_{\mu\nu}K+2g^{55}K_{\mu}^{\lambda}
K_{\lambda \nu}.
\ee
For our metric, $K_{\mu\nu}=g_{\mu\nu}A^\prime/2A$ and $K=2A^\prime/A$.
Integrating over $y$, we obtain the 4-D effective action:
\ba
\label{eff4da}
&& {\bf S}^{\left(4\right)}=\int d^4x \sqrt{-\hat\gamma}
\left[\frac{R^{\left(4\right)}}{2\kappa_N^2} - \Le + \sum_i {\bf L}_i \right]
\ea
where the 4-D gravitational coupling is 
\be
\label{GNdefine}
 \frac{1}{2\kappa_N^2} = \frac{1}{2\kappa^2} \int A(y) dy,
\ee
and the effective 4-D cosmological constant is: 
\ba
\Le &=& \left[-\int \Lambda A(y)^2\, dy \right]
 + \left[ \sum_i A(L_i)^2\sigma_i \right]
 + \left[ \frac{1}{\kappa^2}\int \left( A A^{\prime\prime} 
+ (A^\prime)^2\right)\,dy\right] \\
 &&+ \left[  \frac{1}{\kappa^2}\int \left( A A^{\prime\prime} 
- \frac{1}{2}(A^\prime)^2\right)\,dy\right] 
 + \left[ -\frac{2}{\kappa^2}AA^\prime|_{boundaries} \right].
\ea
The five separate contributions to $\Le$ come from the bulk cosmological 
constant, $\Lambda$; the brane tension $\sigma$; the extrinsic curvature, 
$K_{\mu\nu}$, and its derivatives; 
the diagonal component of the Ricci tensor in the extra 
dimension, $R^{(5)}_{55}$; and the Hartle-Hawking term, for spaces with 
boundary.  An integration by parts simplifies the last three terms, leaving:
\ba
\label{deltaL}
\Le &=& \left[-\int \Lambda A(y)^2\, dy \right]
 + \left[ \sum_i A(L_i)^2\sigma_i \right]
 + \left[ -\frac{3}{2\kappa^2}\int A^\prime(y)^2\,dy \right] \\
&\equiv& \delta_\Lambda + \delta_\sigma + \delta_R .
\ea
For spaces without boundary, the contribution from extrinsic curvature 
vanishes, and the final term, $\delta_R$, arises solely from $R^{(5)}_{55}$.

When $A(y)$ satisfies the Einstein equation (\ref{eq:einstein}),
we can use Eq.(\ref{GNdefine}) and the relation
$H^2 = \kappa_N^2\Le/3$ to obtain: 
\ba
\delta_R &=& -\frac{3}{2\kappa^2}\int A^\prime(y)^2\,dy 
 = -\frac{6H^2}{\kappa^2}\int A(y)\,dy - \int \Lambda A(y)^2\,dy \nonumber \\
 &=& -\frac{6H^2}{\kappa_N^2} -\int \Lambda A(y)^2\,dy 
\label{deltaR}
 = -2\Le -\int \Lambda A(y)^2\,dy.
\ea
Following from Eqs.(\ref{deltaL}),(\ref{deltaR}), we see immediately that if 
$\Le$ is small, then $\delta_\Lambda$ and $\delta_R$ are 
approximately equal and of the same sign, while $\delta_\sigma$ is
twice as large and of the opposite sign. 
If we now use the definitions $q_i\equiv \kappa^2\sigma_i/3$, 
$k_i^2\equiv \kappa^2\Lambda_i/6$, and introduce:
\be
\label{XYdefine}
X \equiv \sum_i q_i A(L_i)^2, \qquad  
Y \equiv \int (2k)^2 A(y)^2\,dy 
\ee
it then follows that the 4-D effective cosmological constant is 
$\kappa^2\Le = X - Y$,
and the separate contributions are:
\be
\kappa^2\delta_\Lambda = -\frac{3}{2}Y,\qquad \kappa^2\delta_\sigma = 3X,
\qquad \kappa^2\delta_R = -2X +\frac{1}{2}Y.
\ee

As a simple illustration, 
consider the case shown in Fig.{\thefigcount}, but where
$q_1=0$ and $k_0=k_1=k_2=k_3$ (i.e., zero brane charges).  
(Here we set $L_0=0$ and $L_2-L_0\equiv L$.)
Using the bulk solution for the scale factor $A(y)$ (\ref{metricA}), 
for $-y_0<y<y_0$, with particle horizons at $\pm y_0$,
the jump conditions are ($\xi \equiv e^{-2ky_0}$):
\ba
\frac{2k}{q_0} &=& \coth[ky_0] = \frac{1-\xi}{1+\xi}\\
\frac{2k}{q_2} &=& \coth[k(L-y_0)] = 
\frac{\xi-e^{-2kL}}{\xi+e^{-2kL}}.
\ea
For large brane separation, $q_0L\gg 1$, these conditions imply that
\ba
k &=& \frac{q_0}{2}\left( 1 - 2\xi + {\cal O}(\xi^2) \right) \\
\xi &=& \frac{q_0 + q_2}{q_2-q_0} e^{-q_0L} + {\cal O}(\xi^2). 
\ea
We next need to calculate the quantities $X$ and $Y$ defined in
Eq.(\ref{XYdefine}):
\ba
X &=& q_0 A(0)^2 \\
Y &=& (2k)^2 A(0)^2 \int_{-y_0}^{y_0} \frac{\sinh^4[k(|y|-y_0)]}{\sinh^4[ky_0]}\,dy \\
&=& 2kA(0)^2 \left[ 1- 4\xi + {\cal O}(\xi^2) \right] \\
&=& q_0 A(0)^2 \left[ 1 - 6\frac{q_0 + q_2}{q_2-q_0}e^{-q_0L} + \cdots \right]
\ea
Note that the $y$ integration is taken over the region between particle 
horizons  which contains the $q_0$ brane (i.e. $-y_0 < y < y_0$). 
We can now list the various contributions to $\Le$:
\ba
\kappa^2\delta_\Lambda = -\frac{3}{2}Y &=& A(0)^2\left[-\frac{3}{2}q_0 + 9q_0\frac{q_0+q_2}{q_2-q_0}e^{-q_0L} + \cdots\right] \\
\kappa^2\delta_\sigma = 3X &=& A(0)^2\left[ 3q_0 \right] \\
\kappa^2\delta_R = -2X + \frac{1}{2}Y &=& A(0)^2\left[ -\frac{3}{2}q_0 - 3q_0\frac{q_0+q_2}{q_2-q_0}e^{-q_0L} + \cdots\right] \\
\kappa^2\Le = \kappa^2\delta_\Lambda + \kappa^2\delta_\sigma 
+ \kappa^2\delta_R 
&=& A(0)^2\left[ 6q_0\frac{q_0+q_2}{q_2-q_0}e^{-q_0L} + \cdots\right]  .
\ea
To relate $\Le$ to $H^2$, we need:
\ba
\frac{\kappa^2}{\kappa_N^2} &=&  \int A(y) 
=  A(0) \int_{-y_0}^{y_0}\frac{\sinh^2[k(|y|-y_0)]}{\sinh^2[ky_0]}\,dy 
= \frac{A(0)}{k} + \cdots 
\ea
Then $H^2 = (\kappa_N^2/\kappa^2)(\kappa^2\Le)/3$, or:
\be
\frac{H^2}{A(0)} = q_0^2\frac{q_0+q_2}{q_2-q_0}e^{-q_0L}.
\ee
This agrees with the appropriate limit of the first of Eqs.(\ref{H_effs}).  
This approach, derived from the 4-D effective action, 
can be used similarly in other brane configurations
to obtain the 4-D effective cosmological constant.  

We thank Horace Stoica for discussions.
This research is partially supported by NSF (E.E.F. and S.-H.H.T.) and NASA
(I.W.).

\end{document}